\documentclass[preprint,showpacs,amsmath,amssymb,aps,prc,superscriptaddress]{revtex4}
\usepackage{epsfig}
\begin{document}

\preprint{APS/123-QED}

\title{\bf \large Binary reaction decays from $^{24}$Mg + $^{12}$C} 

\bigskip

\author{C. Beck\thanks{Correponding author: 
christian.beck@ires.in2p3.fr},$^{1,*}$
P. Papka,$^{1,+}$
A.~S\`{a}nchez i Zafra,$^{1}$
S. Thummerer,$^{1,2}$
F. Azaiez\thanks{IPN Orsay, Orsay, France},$^{1,**}$ 
P. Bednarczyk,$^{1}$ 
S. Courtin,$^{1}$
D. Curien,$^{1}$ 
O. Dorvaux,$^{1}$ 
D. Lebhertz,$^{1}$
A. Nourreddine,$^{1}$ 
M. Rousseau
}

\affiliation{\it Institut Pluridisciplinaire Hubert Curien - D\'epartement de
Recherches Subatomiques, UMR7178, 
IN2P3-CNRS et Universit\'{e} de Strasbourg, 23 rue du Loess, 
B.P. 28, F-67037 Strasbourg Cedex 2, France} 

\medskip

\author{
W. von Oertzen,$^{2}$
B. Gebauer,$^{2}$
C. Wheldon,$^{2.++}$
Tz. Kokalova
}

\address{\it Hahn-Meitner-Institut, Glienicker Str. 100, D-14109 
Berlin, Germany}

\medskip

\author{G.~de~Angelis,$^{3}$
A. Gadea,$^{3}$
S. Lenzi,$^{3}$ 
S. Szilner,$^{3,***}$
D.~R Napoli
}

\address{\it INFN-Lab. Nationali di Legnaro and Dipartimento di 
Fisica, I-35020 Padova, Italy} 

\author{W.~N. Catford}

\address{\it School of Physics and Chemistry, University of Surrey, 
Guildford, Surrey, GU2 7XH, UK} 

\author{D.~G. Jenkins}

\address{\it Department of Physics, University of York, York, 
YO10 5DD, UK} 

\author{G. Royer}

\address{\it Subatech, IN2P3-CNRS et Universit\'e-Ecole des Mines, 
4 rue A. Kastler, F-44307 Nantes Cedex 3, France} 

\date{\today}

\newpage

\begin{abstract}

{Charged particle and $\gamma$-decays in $^{24}$Mg$^{*}$ are investigated for 
excitation energies where quasimolecular resonances appear in $^{12}$C+$^{12}$C 
collisions. Various theoretical predictions for the occurence of superdeformed and 
hyperdeformed bands associated with resonance structures with low spin are 
discussed within the measured $^{24}$Mg$^{*}$ excitation energy region. The 
inverse kinematics reaction $^{24}$Mg$+^{12}$C is studied at E$_{lab}$($^{24}$Mg) 
= 130 MeV, an energy which enables the population of $^{24}$Mg states decaying 
into $^{12}$C+$^{12}$C resonant break-up states. Exclusive data were collected 
with the Binary Reaction Spectrometer in coincidence with {\sc EUROBALL IV} 
installed at the {\sc VIVITRON} Tandem facility at Strasbourg. Specific 
structures with large deformation were selectively populated in binary reactions 
and their associated $\gamma$-decays studied. Coincident events associated with 
inelastic and $\alpha$-transfer channels have been selected by choosing the 
excitation energy or the entry point via the two-body {\it Q}-values. The 
analysis of the binary reaction channels is presented with a particular emphasis 
on $^{24}$Mg-$\gamma$, $^{20}$Ne-$\gamma$ and $^{16}$O-$\gamma$ coincidences.
New information (spin and branching ratios) is deduced on high-energy states in 
$^{24}$Mg and $^{16}$O, respectively.}   

\end{abstract}

\bigskip

%{\bf PACS} number(s): 
\pacs{25.70.Jj, 25.70.Pq, 24.60.Dr; 25.70,+e}

\maketitle

\noindent
$^{*}$ Corresponding author: christian.beck@ires.in2p3.fr \\
$^{+}$ Present address: Department of Physics, University of
Stellenbosch, Matieland, Stellenbosch 7602, South Africa \\
$^{**}$ Permanent address: IPN Orsay, Orsay, France \\
$^{++}$ Present address: School of Physics and Astronomy, University of 
Birmingham, Edgbaston, Birmingham B15 2TT, United Kingdom \\
$^{***}$ Permanent address: Ruder Bo\v{s}kovi\'{c} Institute, HR-10001 Zagreb, 
Croatia \\
\newpage

\section{Introduction}

The observation of resonant structures in the excitation functions for various 
combinations of light $\alpha$-cluster (N = Z) nuclei in the energy regime from 
the barrier up to regions with excitation energies of E$_{x}$ = 20-50~MeV remains 
a subject of contemporary debate~\cite{Greiner95}. These resonances have been 
interpreted for $^{12}$C+$^{12}$C~\cite{Erb85}, the most favorable case for the 
observation of quasimolecular resonances
\cite{Greiner95,Erb85,Morsad91,Beck94,Szilner97}, in terms of nuclear 
molecules. The question whether the well known $^{12}$C+$^{12}$C 
quasimolecular resonances represent true cluster states in the $^{24}$Mg compound 
system, or whether they simply reflect scattering states in the ion-ion potential 
is still unresolved~\cite{Greiner95,Beck04a,Beck04b,Beck04c}. In many cases, 
these resonant structures have been associated with strongly-deformed shapes and 
with clustering phenomena, predicted from the cranked $\alpha$-cluster model
\cite{Marsh86}, the Nilsson-Strutinsky approach~\cite{Leander75,Aberg94}, 
Hartree-Fock calculations~\cite{Flocard84} or other mean-field calculations
\cite{Gupta08}. Of particular interest is the relationship between 
superdeformation (SD) and nuclear molecules, since nuclear shapes with 
major-to-minor axis ratios of 2:1 have the typical ellipsoidal elongation (with 
quadrupole deformation parameter $\beta_2$ $\approx$ 0.6) for light nuclei
\cite{Aberg94}. Furthermore, the structure of possible octupole-unstable 3:1 
nuclear shapes (with $\beta_2$ $\approx$ 0.9) - hyperdeformation (HD) - for 
actinide nuclei has also been widely discussed
\cite{Aberg94,Cwiok94,Cseh04,Andreev06,Adamian07} in terms of clustering 
phenomena. 

Various decay branches of highly excited $^{24}$Mg$^*$ resonant states from 
$^{12}$C+$^{12}$C, including the emission of $\alpha$-particles or heavier 
fragments, are energetically favored. However, $\gamma$-decays have not been 
observed so far, and the $\gamma$-ray branches are predicted to be rather 
small at these excitation energies. Some old experiments have been reported
\cite{McGrath81,Metag82}, which have searched for these very small branching 
ratios expected to be in the range of $10^{-4}~-~10^{-5}$ of the total 
width~\cite{Baye84,Uegaki98,Zhang97}. Very recently the radiative capture 
reaction $^{12}$C+$^{12}$C has been investiugated~\cite{Jenkins07}. The 
rotational bands, from the knowledge of the measured spins and excitation 
energies, can be extended to rather low angular momenta, where finally the 
$\gamma$-decay becomes a larger part of the total decay width. The population 
of such states in $\alpha$-cluster nuclei, which lie below the threshold for 
fission decays and for other particle decays, is favored in binary reactions, 
where at a fixed incident energy the compound nucleus is formed at an excitation 
energy governed by the two-body reaction kinematics. These states may be
coupled to intrinsic states in $^{24}$Mg$^{*}$ populated by a break-up process 
(via resonances) as shown in Refs.\cite{Fulton86,Wilczynski86,Curtis95,Singer00}. 
The $^{24}$Mg+$^{12}$C reaction has been extensively investigated by several 
measurements of the $^{12}$C($^{24}$Mg,$^{12}$C$^{12}$C)$^{12}$C break-up channel
\cite{Fulton86,Curtis95,Singer00}. Resonant breakups are found to 
occur from specific states in $^{24}$Mg at excitation energies E$_{x}$ = 20-40~MeV
(with spins ranging from J = 4$\hbar$ to 14$\hbar$), which are linked to the 
ground state and also have an appreciable overlap with the $^{12}$C+$^{12}$C 
quasimolecular configuration. Several attempts \cite{Curtis95} were made 
to link the $^{12}$C+$^{12}$C barrier resonances \cite{Erb85} with the break-up 
states. The underlying reaction mechanism is now fairly well established
\cite{Singer00} and many of the barrier resonances appear to be correlated, 
indicating that a common structure is involved. This correlation strongly supports 
the hypothesis of the link between barrier resonances~\cite{Erb85,Beck94} 
in $^{12}$C+$^{12}$C, and secondary minima in the Compound Cucleus (CN) $^{24}$Mg 
\cite{Marsh86,Flocard84,Leander75}. 
 
Large quadrupole deformations ($\beta_2$~=~0.6-1.0) and $\alpha$-clustering in 
light N = Z nuclei are known to be general phenomena at low excitation energy. For 
high angular momenta and higher excitation energies, very elongated shapes are 
expected to occur in $\alpha$-like nuclei for A$_{\small CN}$ = 20-60. These 
predictions come from the generalized liquid-drop model, taking into account the 
proximity energy and quasi-molecular shapes~\cite{Royer07}. In fact, highly 
deformed shapes and SD rotational bands have been recently discovered 
in several such N = Z nuclei, in particular, $^{36}$Ar \cite{Svensson00} and 
$^{40}$Ca \cite{Ideguchi01} using $\gamma$-ray spectroscopy techniques. 
Hyperdeformed (HD) bands in $^{36}$Ar nucleus and its related ternary 
clusterizations are theoretically predicted \cite{Algora06}. With the exception 
of the cluster decay of $^{60}$Zn \cite{Zherebchevsky07,Oertzen08a} and $^{56}$Ni 
\cite{Oertzen08b,Oertzen08c,Wheldon08} recently studied using charged particle 
spectroscopy, no evidence for ternary breakup has yet been reported 
\cite{Curtis96,Murphy96,Beck96,Beck98,Sanders99,Gupta08} in light nuclei; the 
particle decay of $^{36}$Ar SD bands (and other highly excited bands) is still 
unexplored. The main binary reaction channels of the $^{24}$Mg+$^{12}$C reaction, 
for which both quasimolecular structures \cite{Beck94,Ford79,Shimizu82} and 
orbiting phenomena \cite{Glaesner90} have been observed, 
is investigated in this work by using charged particle-$\gamma$-ray coincidence 
techniques. 

The present paper is organized in the following way: Sec.~II describes the 
experimental procedures and the data analysis. Sec.~III presents the exclusive 
$^{24}$Mg+$^{12}$C data (part of the experimental results presented here in 
detail have already been shortly reported elsewhere \cite{Beck08,Beck09}) which 
are discussed in Sec.~IV. We end with a summary of the main results and brief 
conclusions in Sec.~V.

\newpage

\section{Experimental set-up}

The study of charged particle-$\gamma$-ray coincidences in binary reactions in 
inverse kinematics is a unique tool in the search for extreme shapes related to 
clustering phenomena. In this paper, we investigate the $^{24}$Mg+$^{12}$C reaction 
with high selectivity at a bombarding energy E$_{lab}$($^{24}$Mg) = 130 MeV 
by using the Binary Reaction Spectrometer (BRS)
\cite{Beck04c,Thummerer00,Gebauer03} in coincidence with the
{\sc EUROBALL IV} (EB)~\cite{Euroball} $\gamma$-ray spectrometer installed at the 
{\sc VIVITRON} Tandem facility of Strasbourg 
\cite{Beck04a,Beck04b,Beck04c,Beck08,Beck09}. The $^{24}$Mg beam was produced and
accelerated by the {\sc VIVITRON}, negative MgH$^{-}$ ions were extracted from the 
ion source and then the MgH molecules  were cracked at the stripping foils of the 
terminal accelerator. The beam intensity was kept constant at approximately 5 pnA.

The targets consisted of 200 $\mu$g/cm$^2$ thick foils of natural $^{12}$C. The 
effects of heavy contaminants (mainly $^{63}$Cu and smaller traces of W isotopes) 
have been estimated to be negligible. The presence of oxygen in the 
$^{12}$C target was not significant (less than $\approx$ 1$\%$). The choice of the 
$^{12}{\rm C}(^{24}{\rm Mg},^{12}{\rm C})^{24}{\rm Mg^{*}}$ reaction implies that 
for an incident beam energy of E$_{lab}$ = 130~MeV an excitation energy range up to 
E$^{*}$ = 30~MeV in $^{24}$Mg is covered~\cite{Curtis95}. 

The {\sc EUROBALL IV} $\gamma$-ray spectrometer, as shown in Fig.~1, consisted of 
two rings of a total of 26 Clover-Ge-detectors each composed of four Ge crystals
\cite{Duchene99} located at angles around 90$^\circ$, and  15 Cluster-Ge-detectors
\cite{Eberth96} each consisting of seven Ge crystals at backward angles. The inner 
BGO scintillator array was removed.

The BRS, in conjunction with EB, gives access to a novel approach for the study of 
nuclei at large deformations as described below.

\subsection{Experimental set-up of the BRS}

\begin{figure}
\caption{\label{fig1} (Color online) Schematic drawing of the scattering chamber 
showing the BRS arrangement for fragment detection and the $\gamma$-ray detectors 
of EB. At forward angles the two BRS gas detector telescopes are depicted, as well 
as two rings of Clover-Ge-detectors at angles around $\theta$ = 90$^\circ$ and 
Cluster-Ge-detectors at backward angles, respectively.}
\begin{center}
\psfig{figure=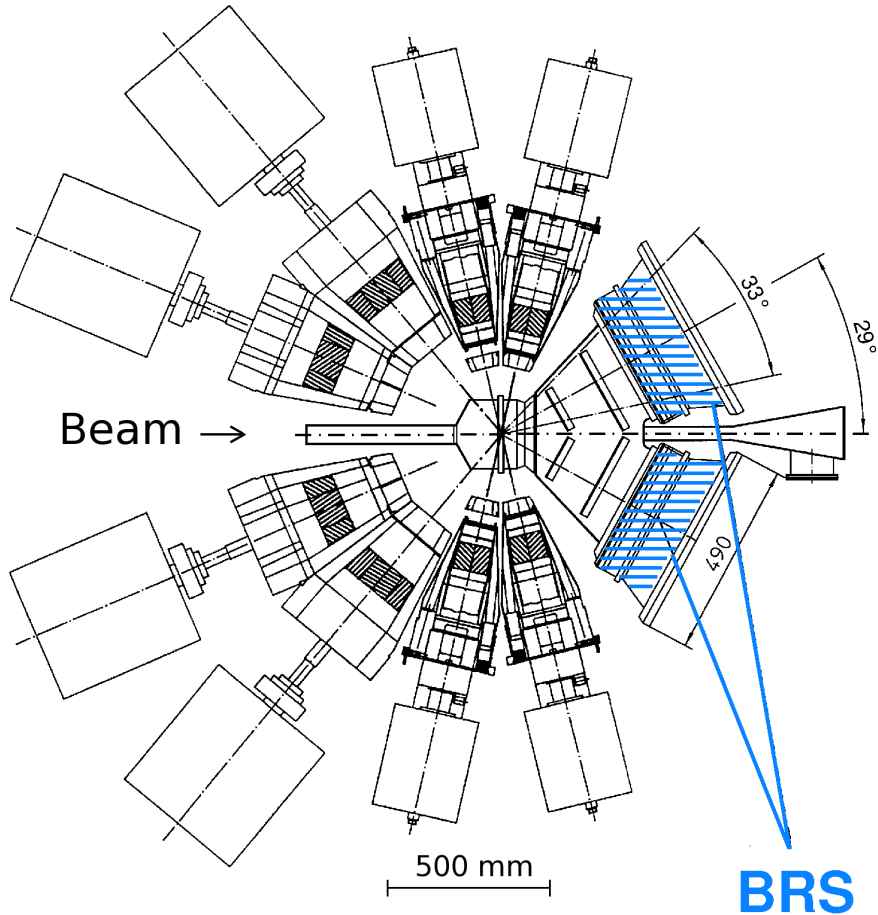,width=12cm,clip=}
\end{center}
\end{figure}

The BRS associated with EB combines as essential elements two large-area (with a 
solid angle of 187 msr each) heavy-ion gas-detector telescopes in a kinematical 
coincidence setup at forward angles. A schematic lay-out of the actual experimental 
set-up of the BRS with EB is shown in Fig.~1. 

A photograph of one of the two BRS telescopes is shown in Fig.~2 (top) along with 
a two-dimensional spectrum (bottom) obtained with a mask during a $^{32}$S+$^{197}$Au 
calibration run at 163.5 MeV \cite{Wheldon08}. The BRS positions were calibrated using 
the elastic scattering data along with the known positions of slits in the movable 
mask in front of the BRS telescope 1.
  
\begin{figure}
\caption{\label{fig2} (Color online) Photograph (top) showing the two sides of the 
mask in place, in front of one arm of the BRS telescope. The corresponding calibrated 
two-bidimensional position $x$ versus $y$ (in mm) spectrum (bottom) has been 
registered using a 210 $\mu$g/cm$^2$ $^{197}$Au target with the 163.5 MeV $^{32}$S 
beam. }
\begin{center}
\psfig{figure=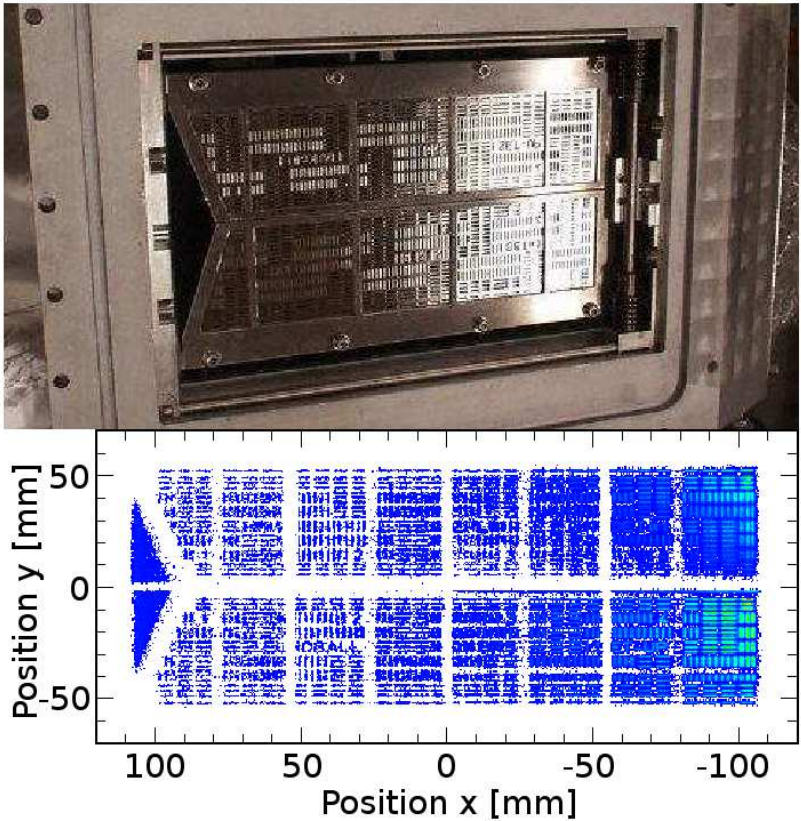,width=11cm,clip=}
\end{center}
\end{figure}
  
The two BRS telescope arms are mounted symmetrically on either side of the beam axis, 
each covering the forward scattering angle range $\Delta$$\theta$ = 
12.5$^\circ$-45.5$^\circ$, i.e. $\theta$ = 29$^\circ$ $\pm$ 16.5$^\circ$. For this 
reason the 30 tapered Clover-Ge-detectors of EB were removed.

\begin{figure}
\caption{\label{fig3} (Color online) Two dimensional BP versus E spectrum, using 
fragment-fragment coincidences, measured in the $^{24}$Mg(130~MeV)+$^{12}$C reaction 
with the BRS. For Z-identification, the Z=12 gate is highlighted to show how the 
$^{24}$Mg events have been selected for the $\gamma$-ray spectrum displayed in 
Fig.~5. The three arrows indicate unambiguously the C, O and Ne fragments. Their
corresponding end-point/punch-through energies, discussed in the text, can be 
clearly seen.}
\begin{center}
\psfig{figure=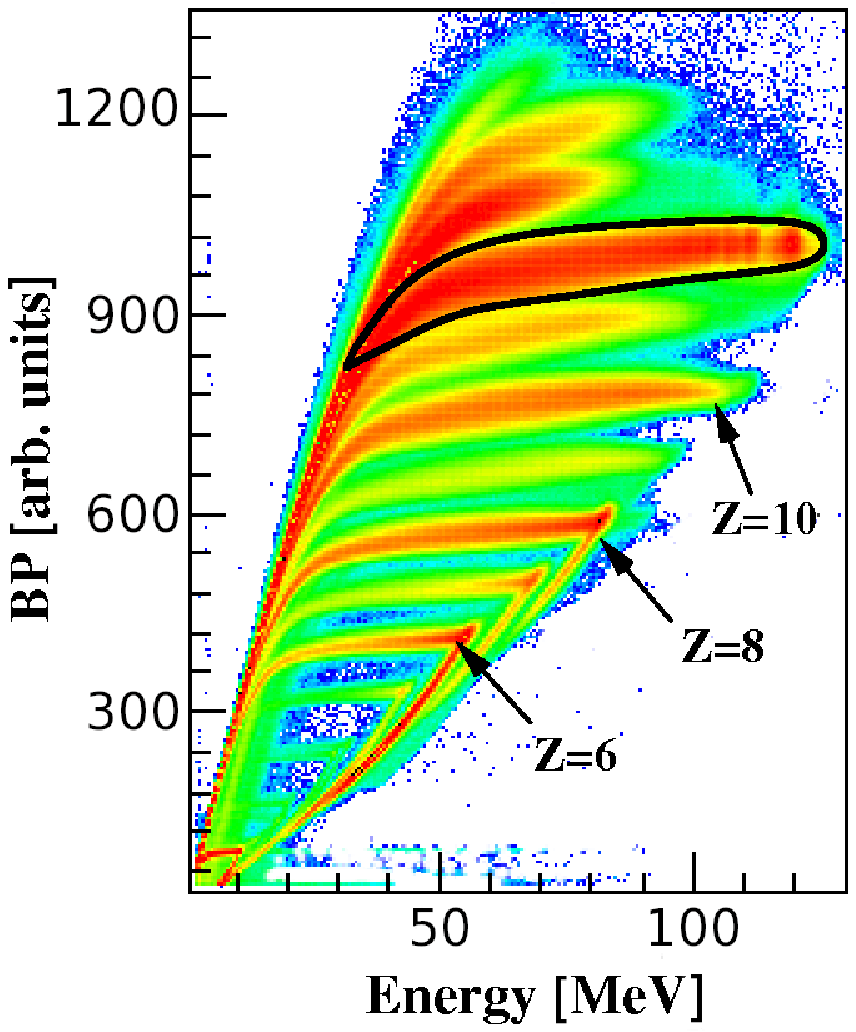,width=9cm,clip=}
\end{center}
\end{figure}

Each BRS gas-detector telescope comprises two consecutive gas volumes containing a 
two-dimensional position sensitive low-pressure multi-wire chamber (MWC) and a 
Bragg-curve ionization chamber (BIC), respectively. All detection planes are 
four-fold segmented in order to improve the resolution and to increase the 
counting rate capability (10$^5$ events/s). The position and time resolutions have
been determined to be intrinsically $\approx$ 0.5 mm and $\approx$ 200 ps, 
respectively \cite{Thummerer00,Gebauer03}. In the MWCs, the in-plane and 
out-of-plane scattering angles $\Theta$ and $\Phi$, respectively, are derived from 
the position ($x$ and $y$) measurements. As already shown in previous studies
\cite{Beck98,Sanders99}, the in-plane angular correlations of two fragments need 
coincidence measurements for binary fission yields to be measured adequately. 
Unfortunately, the MWC of BRS telescope 2 could not provide us with a well 
functionning $y$-position signal (i.e. $\Phi_2$), which is an essential 
information to check kinematical conditions of the out-of-plane correlations of  
ternary fission fragments. In the BICs the Bragg-Peak height (BP), Range (R), and 
kinetic energy (E) are measured. Two-body {\it Q}-value spectra have been reconstructed 
using events for which both fragments are in well selected states. The reaction 
mechanism - projectile breakup or ternary fission - responsible for the 
population of particular states could be determined from two-body kinematics. The 
two heavy fragments are registered using a kinematic coincidence method and 
identified by their charges Z. The excellent elemental identification (Z) is 
illustrated by two-dimensional plots showing BP versus E. Fig.~3 displays a 
typical example of a two-dimensional BP vs E spectrum obtained for the 
$^{24}$Mg+$^{12}$C reaction. Energy calibrations were done using the energy loss 
program SRIM-2003 \cite{Ziegler} (taking into account the window foils and gases) 
to calculate the end-point energies (punch-through) observed in the BP vs E plots. 
Energy spectra are given in Figs.~3 and~4 with absolute energy scales. The 
spectrum of Fig.~3 shows the excellent Z discrimination (the fragments with Z = 6, 
8 and 10 are indicated with respective arrows) achieved with the BICs. It can be 
observed that, due to {\it Q}-value effects, the $\alpha$-like $^{12}$C, $^{16}$O, 
$^{20}$Ne, $^{24}$Mg and $^{28}$Si nuclei are preferentially populated compared 
to the odd Z nuclei. This strong odd-even effect is a characteristic behavior of 
structure phenomena in $^{12}$C+$^{24}$Mg orbiting \cite{Glaesner90} that may 
still survive at E$_{lab}$ = 130 MeV. The gas pressure has been optimized to stop 
ions with Z larger than 9 (ejectiles with lower Z's are not stopped in the BICs 
as shown by their characteristic end point/punch-through in Fig.~3).

The Z=12 gate of Fig.~3 has been used to select the $\gamma$-ray spectrum measured 
in coincidence with $^{24}$Mg nucleus (see Fig.~5). Other well-defined Z gates can 
be used for the processing of the $\gamma$-ray spectra in coincidence with the 
$^{12}$C, $^{16}$O, $^{20}$Ne and $^{28}$Si nuclei of interest. Although mass 
identification has not been available in this experiment - i.e. the performance
of the VIVITRON pulsing system was not better than 2ns (FWHM) -, the good energy 
resolution along with Z-identification allowed the different reaction mechanisms
to be clearly discriminated. 

\begin{figure}
\caption{\label{fig4} (Color online) Two-dimensional angle versus energy spectrum, 
using fragment-fragment coincidences, measured for the $^{16}$O+$^{20}$Ne 
exit-channel. The relative intensity is shown on the side bar. The dashed red line 
corresponds to the high-energy cutoff due to the geometrical bias of the kinematical 
coincidences. The regions labelled (a) to (f) are defined in detail in the text. They 
correspond to the following excitation energies ranges E$^{*}$ = 0-3 MeV (a), 
3-6 MeV (b), 6-10 MeV (c), 10-14 MeV (d), and 14-19 MeV (e), and finally region [VI] 
is defined for E$^{*}$ larger than 19 MeV (f).}
\begin{center}
\psfig{figure=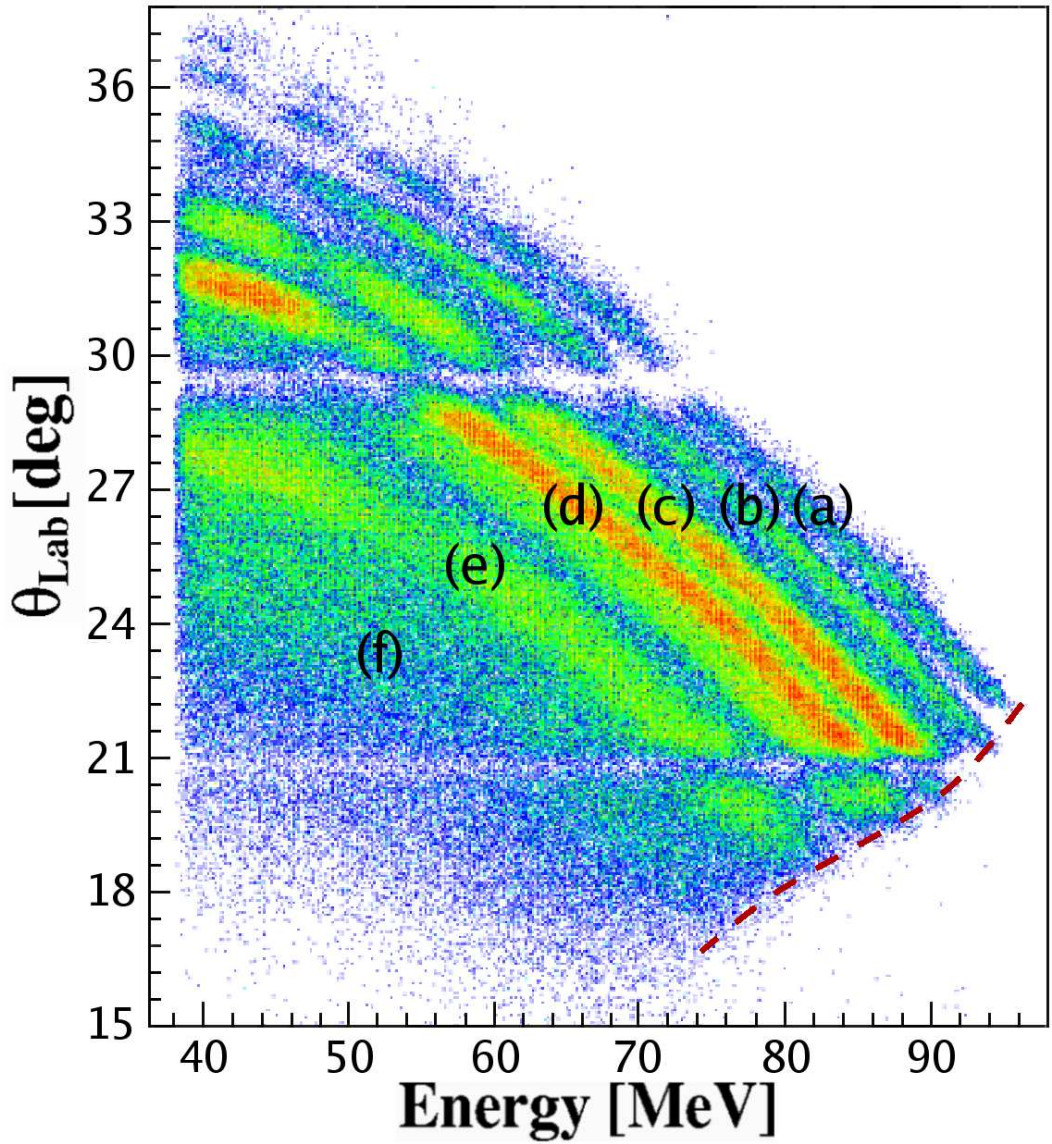,width=14cm,clip=}
\end{center}
\end{figure}

Fig.~4 illustrates a typical example of a two-dimensional angle versus energy 
spectrum for the $^{16}$O+$^{20}$Ne exit-channel. The six regions (a-f) labelled [I] to
[VI] have been defined as a function of the inelasticity of the reaction channel
from the ground-state {\it Q}-value E$^{*}$ = 0 to full damping with E$^{*}$ larger
than 15 MeV. This $\alpha$-transfer channel will be discussed in Sec.~III.B. 
It should be noticed that the two lines around 21$^{\circ}$ and 29$^{\circ}$ 
correspond to shadows produced by the supports to Mylar entrance windows of the 
MWC of the BRS telescope 1. The dashed line corresponds to the high-energy cutoff 
due to the geometrical bias of the kinematical coincidences. More details of the 
detectors and the experimental set-up of the BRS are given in Refs.
\cite{Beck04c,Beck08,Beck09,Thummerer00,Gebauer03}. 

\subsection{Experimental set-up of EUROBALL IV}

The excellent channel selection capability of binary and/or ternary fragments (see 
for instance the two-dimensional angle versus energy spectrum plotted in Fig.~4 
for the $\alpha$-transfer channel) allows clear discrimination between the 
reaction channels, so that EB \cite{Euroball} is used mostly with one- or 
two-fold multiplicities, for which the total $\gamma$-ray efficiency is very high. 
The removal of the conventional 30 tapered Clover-Ge-detectors of EB from the 
forward angles had only a very small impact on reducing the overall EB efficiency
(less than 13$\%$ of the total $\gamma$-ray efficiency). Although no specific
measurements were done during two BRS campaigns, the absolute total $\gamma$-ray 
efficiency has been estimated to be approximately 8.5$\%$ for the present EB 
experiment. 
 
In binary exit-channels, the exclusive detection of both ejectiles allows precise 
{\it Q}-value determination and Z-resolution, as shown in Figs.~5 and 6 for 
$^{24}$Mg+$^{12}$C inelastic scattering channel, and the $^{16}$O+$^{20}$Ne 
$\alpha$-transfer channel, respectively. The $\gamma$ rays were Doppler-shift
corrected for the velocity of the identified outgoing binary fragments. The 
Doppler-shift corrections yield an energy resolution better than 40-59~keV (FWHM) 
for 3-4~MeV $\gamma$ rays ( $\approx$ 1.3 $\%$ in average). The $\gamma$-ray 
spectra of Figs.~5 and 6 will be discussed in detail in the next section.

\begin{figure}
\caption{\label{fig5}Doppler corrected $\gamma$-ray spectrum for $^{24}$Mg, using 
fragment-$\gamma$ coincidences, measured with the BRS/EB detection system 
of Fig.~1. The $^{24}$Mg nuclei were selected with the Z=12 gate marked in Fig.~3.
The full details of the identified $\gamma$-ray transitions are given in Table I.}
\begin{center}
\psfig{figure=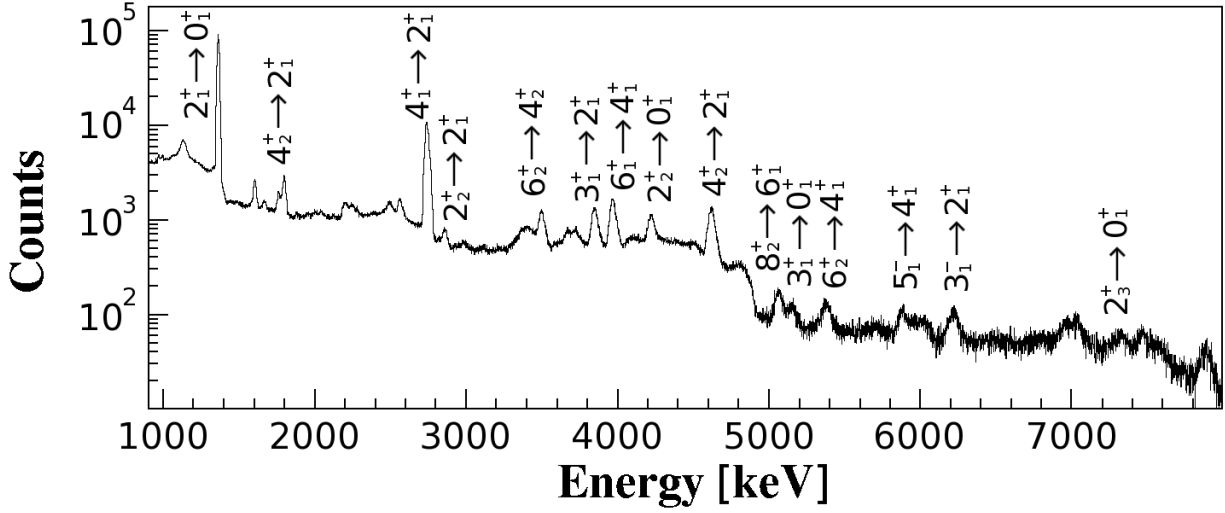,width=16.5cm,clip=}
\end{center}
\end{figure}

\begin{figure}
\caption{\label{fig6} Gated $\gamma$-ray spectra, using 
fragment-fragment-$\gamma$-ray coincidences, measured for the $^{16}$O+$^{20}$Ne 
exit-channel. Doppler-shift corrections were applied for the velocity of the detected 
$^{20}$Ne. The six excitation energy gates labelled (a) to (f) in Fig.~4 
(corresponding to excitation energies ranges E$^{*}$ $\approx$ 0-3 MeV (a), 
3-6 MeV (b), 6-10 MeV (c), 10-14 MeV (d), and 14-19 MeV (e), and finally the 
last gate (f) is applied for E$^{*}$ larger than 19 MeV) have been used as 
triggers to the six $\gamma$-ray spectra. The main $\gamma$-ray transitions in 
$^{20}$Ne are labelled.}
\begin{center}
\psfig{figure=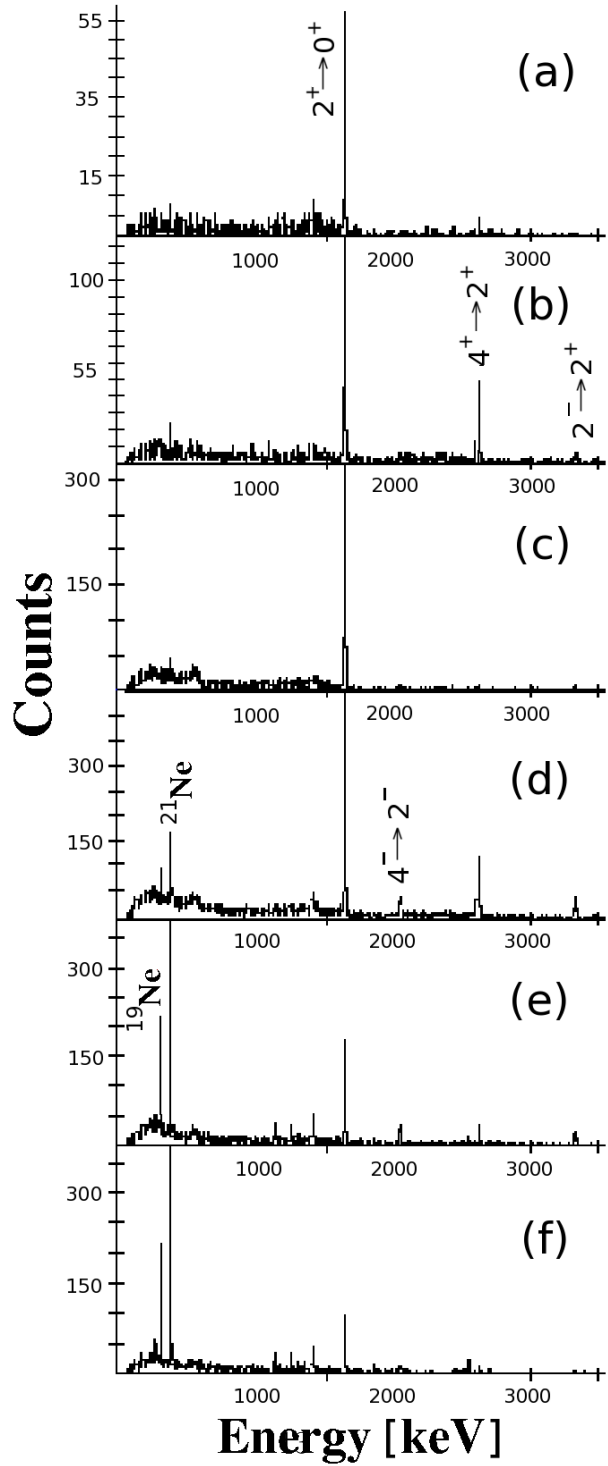,width=8cm,clip=}
\end{center}
\end{figure}

\newpage

\section{Experimental results}

The inverse kinematics of the $^{24}$Mg+$^{12}$C reaction and the negative 
{\it Q}-values give ideal conditions for triggering on the BRS, because the angular 
range is optimum for $\theta_{lab}$ = 12$^\circ$-40$^\circ$ in the lab-system 
(with the range of $\theta_{lab}$ = 12$^\circ$-25$^\circ$ for the recoils), and 
because the solid angle transformation favors the detection efficiency of 
the heavy fragments. Thus we have been able to cover a large part of the angular 
distribution of the binary process with good efficiency. In binary exit-channels, 
the exclusive detection of both ejectiles with precise Z discrimination allows 
excellent Doppler-shift corrections. 

\subsection{$^{24}$Mg+$^{12}$C exit channel}

\begin{table}[c]
\begin{center}
\hspace*{-1cm}
\begin{tabular}{|c|c|c|c|}
%\hline
 E$^{exp}_{\gamma}$(keV) & $\gamma$-ray transition\qquad\qquad\qquad &
E$^{*}_{level}$(keV)\qquad\qquad\qquad& I$_{\gamma}$($\%$)\qquad\qquad\qquad\\
\hline
1368.6 & 2$^{+}_{1}$ $\rightarrow$ 0$^{+}_{1}$\qquad\qquad\qquad & 1368.6\qquad\qquad\qquad & 100 \qquad\qquad\qquad\\
1683.3 & 4$^{-}_{1}$ $\rightarrow$ 3$^{-}_{1}$\qquad\qquad\qquad & 9299.8\qquad\qquad\qquad & 0.32\qquad\qquad\qquad\\
1771.9 & 4$^{+}_{2}$ $\rightarrow$ 2$^{+}_{2}$\qquad\qquad\qquad & 6010.3\qquad\qquad\qquad & 2.11\qquad\qquad\qquad\\
2576.9 & 5$^{+}_{1}$ $\rightarrow$ 3$^{+}_{1}$\qquad\qquad\qquad & 7812.2\qquad\qquad\qquad & 2.11 \qquad\qquad\qquad\\
2754.0 & 4$^{+}_{1}$ $\rightarrow$ 2$^{+}_{1}$\qquad\qquad\qquad & 4122.9\qquad\qquad\qquad & 39.3\qquad\qquad\qquad\\
2869.5 & 2$^{+}_{2}$ $\rightarrow$ 2$^{+}_{1}$\qquad\qquad\qquad & 4238.4\qquad\qquad\qquad & 2.98\qquad\qquad\qquad\\
3517.0 & 6$^{+}_{2}$ $\rightarrow$ 4$^{+}_{2}$\qquad\qquad\qquad & 9528.0\qquad\qquad\qquad & 4.50\qquad\qquad\qquad\\
3689.0 & (4$^{-},5^{+})$ $\rightarrow$ 4$^{+}_{1}$\qquad\qquad\qquad&7812.2\qquad\qquad\qquad&1.53\qquad\qquad\qquad\\
3747.0 & (6$^{+}$,7$^{-}$,8$^{+}$)$\rightarrow$6$^{+}_{1}$\qquad\qquad\qquad&11860\qquad\qquad\qquad&1.47\qquad\qquad\qquad\\
3866.2 & 3$^{+}_{1}$ $\rightarrow$ 2$^{+}_{1}$\qquad\qquad\qquad & 5235.2\qquad\qquad\qquad & 7.22\qquad\qquad\qquad\\
3990.0 & 6$^{+}_{1}$ $\rightarrow$ 4$^{+}_{1}$\qquad\qquad\qquad & 8113.0\qquad\qquad\qquad & 6.52\qquad\qquad\qquad\\
4238.4 & 2$^{+}_{2}$ $\rightarrow$ 0$^{+}_{1}$\qquad\qquad\qquad & 4238.4\qquad\qquad\qquad & 4.96\qquad\qquad\qquad\\
4641.2 & 4$^{+}_{2}$ $\rightarrow$ 2$^{+}_{1}$\qquad\qquad\qquad & 6010.3\qquad\qquad\qquad & 10.2\qquad\qquad\qquad\\
5063.2 & 0$^{+}_{2}$ $\rightarrow$ 2$^{+}_{1}$\qquad\qquad\qquad & 6432.5\qquad\qquad\qquad & 1.48\qquad\qquad\qquad\\
5099.0 & 8$^{+}_{2}$ $\rightarrow$ 6$^{+}_{1}$\qquad\qquad\qquad & 13213 \qquad\qquad\qquad & 3.17\qquad\qquad\qquad\\
5404.0 & 6$^{+}_{2}$ $\rightarrow$ 4$^{+}_{1}$\qquad\qquad\qquad & 9528.0\qquad\qquad\qquad & 1.13\qquad\qquad\qquad\\
5904.2 & 5$^{-}_{1}$ $\rightarrow$ 4$^{+}_{1}$\qquad\qquad\qquad &10027.9\qquad\qquad\qquad & 1.29\qquad\qquad\qquad\\
6246.9 & 3$^{-}_{1}$ $\rightarrow$ 2$^{+}_{1}$\qquad\qquad\qquad & 7616.5\qquad\qquad\qquad & 1.23\qquad\qquad\qquad\\
6988.3 & 3$^{-}_{2}$ $\rightarrow$ 2$^{+}_{1}$\qquad\qquad\qquad & 8358.1\qquad\qquad\qquad & 0.25\qquad\qquad\qquad\\
7069.5 & 4$^{+}_{3}$ $\rightarrow$ 2$^{+}_{1}$\qquad\qquad\qquad & 8439.3\qquad\qquad\qquad & 0.24\qquad\qquad\qquad\\
7347.8 & 2$^{+}_{3}$ $\rightarrow$ 0$^{+}_{1}$\qquad\qquad\qquad & 7349.1\qquad\qquad\qquad & 0.89\qquad\qquad\qquad\\
7554.0 & 1$^{-}_{1}$ $\rightarrow$ 0$^{+}_{1}$\qquad\qquad\qquad & 7555.3\qquad\qquad\qquad & 0.63\qquad\qquad\qquad\\
7615.2 & 3$^{-}_{1}$ $\rightarrow$ 0$^{+}_{1}$\qquad\qquad\qquad & 7616.5\qquad\qquad\qquad & 0.10\qquad\qquad\qquad\\
7914.3 & 2$^{+}_{4}$ $\rightarrow$ 2$^{+}_{1}$\qquad\qquad\qquad & 9284.4\qquad\qquad\qquad & 0.27\qquad\qquad\qquad\\
8436.8 & 1$^{-}_{2}$ $\rightarrow$ 0$^{+}_{1}$\qquad\qquad\qquad & 8438.4\qquad\qquad\qquad & 0.039\qquad\qquad\qquad\\
8990.2 & 2$^{+}_{5}$ $\rightarrow$ 2$^{+}_{1}$\qquad\qquad\qquad &10360.7\qquad\qquad\qquad & 0.075\qquad\qquad\qquad\\
9816.5 & (1,2,3) $\rightarrow$ 2$^{+}_{1}$\qquad\qquad\qquad &11187.3\qquad\qquad\qquad & 0.023\qquad\qquad\qquad\\

\end{tabular}

\caption{\label{TABLE I:} {\small \sl $\gamma$-ray transitions observed in coincidence 
(see Fig.~5) with $^{24}$Mg (Z=12 gate marked in Fig.~3). The efficiency corrected 
$\gamma$-ray yields (given by their relative intensities), normalized to the 
2$^{+}_{1}$ $\rightarrow$ 0$^{+}_{1}$ 1368.6 keV transition, are given in the last 
column. E$^{*}_{level}$ is the energy of the initial level.}} 

\end{center}
\end{table}

Fig.~5 shows the Doppler-shift corrected $\gamma$-ray spectrum for $^{24}$Mg 
events in coincidence with the Z=12 gate, defined in the BP vs E spectrum of Fig.~3. 
Most of the known transitions of $^{24}$Mg
\cite{Vermeer88,Endt93,NNDC,Beck01,Wiedenhover01,Jenkins05,Salsac08} can be 
identified in the energy range depicted in the figure and their properties are given 
in Table I. Weaker transitions in $^{23}$Mg and $^{25}$Mg (not marked in Fig.~5), 
also selected by the Z=12 gate, correspond to one-neutron transfer processes.
The excitation of the 2$^{+}$, 1808.7 keV state of $^{26}$Mg arises either from
a 2-neutron-transfer process or from a fusion-evaporation mechanism. These transitions 
are not reported in Table I. However, we note, in particular, a doublet visible in 
Fig.~5 near 1600 KeV which is due to the $^{25}$Mg 7/2$^+$ $\rightarrow$ 3/2$^+$ and 
$^{23}$Mg 7/2$^+$ $\rightarrow$ 5/2$^+$ transitions. The broad bump at 3.40 MeV 
corresponds to the 9/2$^+$ $\rightarrow$ 5/2$^+$ $\gamma$-ray transition.

The two most prominent $^{24}$Mg lines correspond, as expected, to the $^{24}$Mg 
transitions 2$^{+}_{1}$ $\rightarrow$ 0$^{+}_{1}$ with E$_{\gamma}$~=~1369~keV 
(100$\%$ relative intensity) and 4$^{+}_{1}$ $\rightarrow$ 2$^{+}_{1}$ with 
E$_{\gamma}$~=~2754~keV (39 $\%$), where 0$^{+}_{1}$, 2$^{+}_{1}$ and 4$^{+}_{1}$ 
are the first members of the $^{24}$Mg K$^\pi$~=~0$^+$ ground state band. At higher 
energies, weaker lines (with relative intensities of 5-10$\%$) are observed, 
corresponding to transitions with E$_\gamma$~=~3990~keV (6$^+_{1}$ $\rightarrow$ 
4$^+_{1}$), 3866~keV (3$^+_{1}$ $\rightarrow$ 2$^+_{1}$), 4238~keV (2$^+_{2}$ 
$\rightarrow$ 0$^+_{1}$) and 4641~keV (4$^+_{2}$ $\rightarrow$ 2$^+_{1}$). The 
6$^{+}_{1}$ level belongs to the ground state band and the 2$^+_{2}$, 3$^+_{1}$ 
and 4$^+_{2}$ to the $^{24}$Mg K$^\pi$~=~2$^+$ band. As expected we see decays 
feeding the yrast line of $^{24}$Mg up to the 8$^{+}_{2}$ level at 13.21~MeV 
(as much as 3 $\%$). The relative intensities of the $\gamma$-ray transitions
given in the last column of Table I are normalized to the 2$^{+}_{1}$ $\rightarrow$ 
0$^{+}_{1}$ 1368.6 keV transition. The $\gamma$-ray spectrum shows mainly strong 
$\gamma$-ray transitions between positive-parity states, however rather intense 
transitions are also observed from negative-parity levels (3$^{-}_{1}$ and 
5$^{-}_{1}$). The 1683~keV (4$^{-}_{1}$ $\rightarrow$ 3$^{-}_{1}$) transition is 
weaker, with a relative intensity of much less than 1$\%$. The deformation 
properties of the $^{24}$Mg nucleus are further discussed in Sec.~IV.A.

\subsection{$^{20}$Ne+$^{16}$O exit channel}

Figs.~6 and 7 display the Doppler-shift corrected $\gamma$-ray spectra for events in 
coincidence with two Z=8 and Z=10 gates (as the one shown for Z=12 in the BP vs E 
spectrum of Fig.~3) defined in both arms of the BRS. Most of the known transitions 
of both $^{16}$O and $^{20}$Ne can be identified in the energy range depicted. The 
six different excitation energy gates displayed in Fig.~4 are used to generate the 
$\gamma$-ray spectra shown in Fig.~6 (low-energy transitions) and discussed in 
Sec.~IV.B. The $\gamma$-ray spectrum (high-energy transitions) of Fig.~7 was 
triggered with the use of the gate labelled (d) for E$^{*}$ $\geq$ 6.5 MeV.

Identifications of the most intense $\gamma$ rays in $^{20}$Ne is straightforward 
and their labelling are given in Fig.~6. As expected for $\alpha$-transfer processes
involving mutual excitations of the binary fragments, we observe decays feeding of 
the yrast line of the $^{20}$Ne nucleus. Two previously unobserved transitions in 
$^{16}$O from the decay of the 3$^{+}$ state at 11.08~MeV, clearly visible in the 
$\gamma$-ray spectrum of Fig.~7, have been identified for the first time (see Fig.~8 
for new partial level scheme). We note that, thanks to the excellent resolving power
of the EB+BRS set-up, the respective first escape peaks positions (as indicated
by the two arrows) of the  6.13 MeV, 6.92 MeV and 7.12 MeV $\gamma$-ray transitions 
in $^{16}$O are also apparent in this spectrum. 
 
\begin{figure}
\caption{\label{fig7} $^{16}$O high-energy $\gamma$-ray specrum produced by the 
$^{16}$O+$^{20}$Ne exit-channel with the gate (d) of Fig.~4 (for E$^{*}$ = 
10-14 MeV as defined in the text). Doppler-shift corrections have been applied 
for O fragments detected in the BRS. The two arrows show the respective first escape 
peaks positions of the 6.13 MeV and 6.92 MeV $\gamma$-ray transitions in $^{16}$O. }
\begin{center}
\psfig{figure=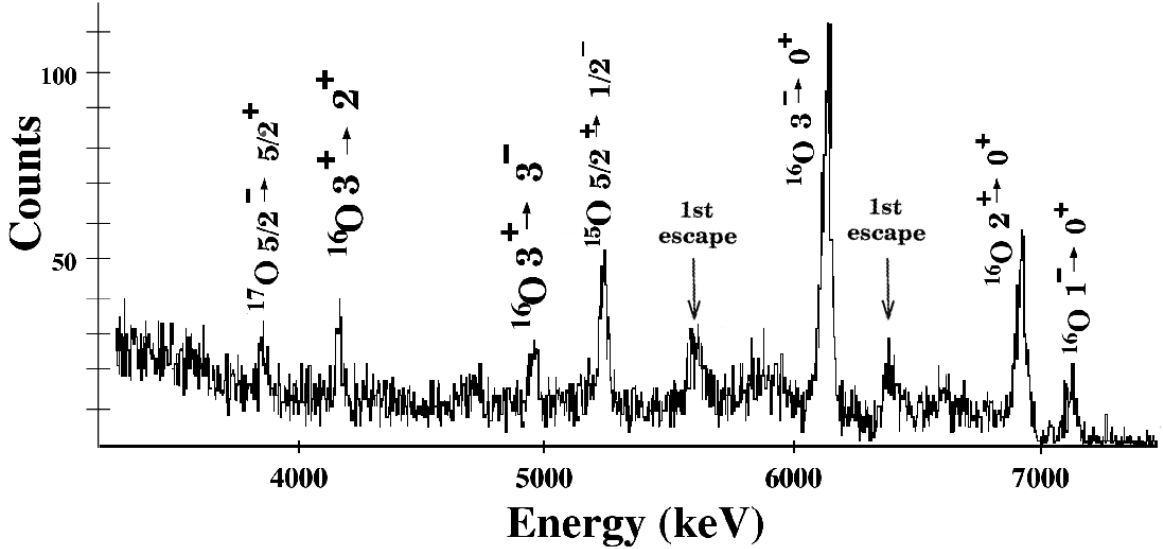,width=15.5cm,clip=}
\end{center}
\end{figure}

The next step in the analysis is the use of the BRS trigger in order to select the 
excitation energy range by the two-body {\it Q}-value (in the $^{16}$O+$^{20}$Ne channel), 
and thus allowing us to study the region at rather low excitation energies (below
and around particle thresholds), where $\gamma$ decay becomes observable. The 
two-dimensional angle versus energy spectrum, shown in Fig.~4 for the 
$\alpha$-transfer channel, allow us to select well defined excitation energy 
({\it Q}-value) regions. We define six different energy regions ranging from the (g.s.) 
elastic and inelastic ($^{20}$Ne, 2$^{+}_1$) transfer (defined as gate (a)) up to 
highly excited states (namely the deep-inelastic gates (e) and (f)). The other
gates correspond to energy regions where mutual inelastic excitation processes 
are dominant. The oscillations observed at large angles in the ``gate [I]" region 
correpond - as for the $^{24}$Mg+$^{12}$C exit-channel (not shown) - to the 
oscillatory nature of the angular distribution of the $\alpha$ transfer, and 
indicate the dominance of the grazing partial waves already observed for the 
$^{24}$Mg+$^{12}$C reaction \cite{Greiner95,Beck94,Ford79,Shimizu82}. The 
second (gate (b)) and third (gate (c)) E$^{*}$ regions include mainly the second 
excited state of $^{20}$Ne (4$^{+}_1$, 4248~keV). The following region (gate (d)) 
with the largest intensity arises from the corresponding mutual excitations with 
the excitation of the (3$^{-}$, 6129~keV) collective state of $^{16}$O. The last 
two gates correspond to more and more energy damped collisions (deep-inelastic, 
orbiting \cite{Glaesner90} and fusion-fission~\cite{Sanders99}) with 
non-oscillatory angular distributions.

We have also analysed the $^{12}$C+$^{12}$C coincidences to investigate
$^{12}$C-$^{12}$C-$^{12}$C ternary decays of $^{36}$Ar. Unfortunately, an
accurate out-of-plane correlation could not measured. Therefore, the hypothesis 
of $^{36}$Ar hyperdeformed shapes, predicted by preliminary calculations performed 
in the framework of the generalized liquid-drop model \cite{Royer07}, will 
need further experimental investigations with higher selectivity. 

\newpage

\section{Discussion}

\subsection{Deformation properties of the $^{24}$Mg nucleus}

The fact that Fig.~5 shows almost exclusively $^{24}$Mg $\gamma$-ray transitions 
between positive-parity states, with weaker transitions observed from 
negative-parity levels (as indicated by Table I) has been explained by the specific 
nature of $^{24}$Mg ~\cite{Beck01} which, in its ground state, has a prolate shape. 
The $^{24}$Mg nucleus appears to be populated primarily through its first two 
rotational bands, K$^{\pi}$~=~0$^{+}$, (2$^{+}_{1}$ (1369~keV), 4$^{+}_{1}$ 
(4123~KeV), 6$^{+}_{1}$ (8113~keV)) and K$^{\pi}$~=~2$^{+}$ (2$^{+}_{2}$ (4236~keV), 
3$^{+}_{1}$ (5235~keV), 4$^{+}_{2} $ (6010~keV), 6$^{+}_{2}$ (9528~keV)), which are 
associated with stable prolate deformations, according to standard shell-model 
calculations \cite{Carchidi88}. In a prior measurement at a lower bombarding energy,
a selective population of natural-parity states has been found to be favored in the 
$^{24}$Mg+$^{12}$C orbiting reaction~\cite{Glaesner90}, with a suppression of the 
3$^{+}_{1}$ (5235~keV) and 5$^{+}_{1}$ (7812~keV) states of the K~=~2 band of 
$^{24}$Mg. In the present data, the 3$^{+}$ state is strongly populated, whereas the 
5$^{+}$ state is moderately populated in sharp contrast with the previous 
experiment~\cite{Glaesner90}. Decays from even higher-energy 6$^{+}_{1}$ (8113~keV) 
and 6$^{+}_{2}$ (9528~keV) levels from the K$^{\pi}$~=~0$^{+}$ and K$^{\pi}$~=~2$^{+}$ 
bands, respectively, are also clearly visible in the spectrum. The selectivity of 
natural parity states in the $^{24}$Mg+$^{12}$C exit-channel indicated that the 
orbiting processes, found to have strong yields at 3-4 times the Coulomb barrier 
\cite{Glaesner90}, are competitive. The rather strong populations of the 3$^{+}$ and 
5$^{+}$ states at E$_{lab}$ = 130 MeV suggest that the orbiting process has a smaller 
contribution than at lower incident energies. This would be in conflict with the 
survival of orbiting-like yields recently measured in the $^{20}$Ne+$^{12}$C reaction 
up to E$_{lab}$ = 200 MeV \cite{Bhattacharya05,Bhattacharya07}.

Although there is an indication of an unknown $\gamma$ ray around 5.95~MeV we cannot 
confirm its interpretation as to be the earlier reported 10$^{+}_{1}$ $\rightarrow$ 
8$^{+}_{2}$~5927 keV transition \cite{Wiedenhover01}.  

The strong population of a broad peak (most probably due to a number of different 
states) around E$_x$ = 10 MeV and states in the non-yrast K~=~2 rotational band was 
observed in the $^{12}$C($^{12}$C,$\gamma$) radiative capture reaction 
\cite{Jenkins05,Jenkins07} at two resonant energies (E$_{c.m.}$ = 6.0 and 6.8 MeV). 
On the other hand the population of the observed states which are members of the 
K~=~2, does appear to be selectively enhanced in the present experiment. 
 
It has been shown using the $\gamma$-$\gamma$ coincidences that most of the states 
of Fig.~5 belong to cascades which contain the characteristic 1369~keV 
$\gamma$-ray and pass through the lowest 2$^{+}$ state in $^{24}$Mg. Still, a 
number of transitions in the high-energy part of the spectrum (6~MeV~-~9~MeV) 
have not been clearly identified. The reason why the search for a direct $\gamma$ 
decay in $^{12}$C+$^{12}$C has not been conclusive so far~\cite{McGrath81,Metag82} 
is due to the excitation energy in $^{24}$Mg as well as the spin region 
(8$\hbar$~-~12$\hbar$) which were probably chosen too high, as in radiative capture 
experiments~\cite{Jenkins05,Jenkins07}, for instance. As the situation about 
$^{12}$C+$^{12}$C quasimolecular resonances is rather complex, future experimental 
works will have to investigate 10~MeV transitions involving low spin (0 or 
2$\hbar$) states. In the case of the break-up reaction, the 4$\hbar$~-~6$\hbar$ 
spin region has been reached, however because of poor statistics none of the 
unplaced transitions can be assigned to the intra-band E$_{2}$ $\gamma$-ray 
expected to be enhanced between resonant molecular states
\cite{Baye84,Uegaki98,Zhang97}.

Although charged-particle decays are strongly competitive at high excitation energy,
high-energy $\gamma$-ray transitions in $^{24}$Mg have been found (but with no clear 
assignement). However, the analysis of the $^{24}$Mg+$^{12}$C exit-channel appears 
inconclusive as far as the search for the intra-band $\gamma$-rays is concerned. One 
should keep in mind that the de-excitation spectra can be very different from the 
excitation ones. Some degrees of freedom can have a major role in the excitation 
mechanism (in matrix elements) whereas they can play a minor one in the de-excitation 
processes, as the de-excitation process may, in fact, be dominated by very small 
components in the wave function. Even if they are strongly excited by some open 
channels (excitation), they can play a very small role in the actual $\gamma$-ray 
spectra we see (de-excitation). Since this intra-band $\gamma$ ray lies so far from 
the $^{24}$Mg yrast line it is much more difficult to identify this band 
experimentally than it was for the analoguous SD bands in heavier nuclei such as 
$^{36}$Ar \cite{Svensson00} and $^{40}$Ca \cite{Ideguchi01}, for instance. In 
addition, the actual branching ratios may be smaller than the calculations using the 
$\alpha$-cluster model \cite{Zhang97}.

\subsection{Gamma-ray spectroscopy of the $^{20}$Ne and $^{16}$O nuclei}

The first two gates of Fig.~4 with excitation energies of 0 $\leq$ E$^{*}$ $\leq$ 
3~MeV and 3 $\leq$ E$^{*}$ $\leq$ 6~MeV, respectively, allow essentially the feeding 
of the first two excited states of the $^{20}$Ne yrast g.s. band to be observed, i.e.
the $\gamma$ rays from 2$^{+}$ $\rightarrow$ 0$^{+}$ and from 
4$^{+}$ $\rightarrow$ 2$^{+}$ are dominant in Fig.~6. When E$^{*}$ is increased by 
choosing the two E$^{*}$ gates: 10 $\leq$ E$^{*}$ $\leq$ 14~MeV and E$^{*}$ $\geq$ 
14~MeV, higher-energy states are observed in $^{20}$Ne. Other $\gamma$ rays from 
$^{21}$Ne and $^{19}$Ne nuclei appear with increasing yields. They correspond 
respectively to the 2{\it p}{\it 1n} multi-nucleon transfer channel (or $\alpha$ 
transfer followed by a subsequent neutron evaporation) and $\alpha${\it n} transfer 
channel, respectively. The 5/2$^{+}$ $\rightarrow$ 1/2$^{-}$ transition of $^{15}$O 
corresponds also to a multi-nucleon transfer (i.e. 2{\it p}1{\it n}) process. It 
is also worth noting that feeding of the $^{20}$Ne states appears to saturate with 
increasing E$^{*}$, since binary decays decrease compared to sequential three-body 
decay channels such as $^{16}$O+$^{20}$Ne$^{*}$ $\rightarrow$ 
$^{16}$O+$^{16}$O+$\alpha$. This saturation effect is known to occur for orbiting 
processes due to an angular momentum limitation as evidenced, for example, in 
$^{28}$Si+$^{12}$C deep-inelastic collisions \cite{Shapira84}.

\begin{figure}
\caption{\label{fig8} New partial (high-energy) level scheme of $^{16}$O 
corresponding to $\gamma$-ray transitions observed in Fig.~7.}
\begin{center}
\psfig{figure=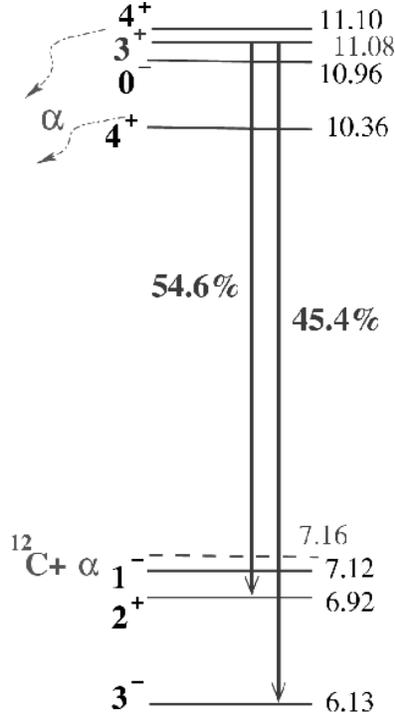,width=5.5cm,clip=}
\end{center}
\end{figure}

With appropriate Doppler-shift corrections applied to Oxygen fragments identified 
in the BRS, it has been possible to extend the knowledge of the level scheme of 
$^{16}$O at high energies~\cite{Endt93,NNDC,Bromley59}, well above the 
$^{12}$C+$\alpha$ threshold (7.162~MeV), which is given in Fig.~8 (dashed line) for 
the sake of comparison. 

New information has been deduced on branching ratios of the decay of the 3$^{+}$ 
state of $^{16}$O at 11.085~MeV $\pm$ 3 keV. This state does not $\alpha$-decay 
because of non-natural parity \cite{Bromley59}, in contrast to the two neighbouring 
4$^{+}$ states (at 10.36~MeV and 11.10~MeV, respectively as indicated in Fig.~8) to 
the 2$^{+}$ state at 6.92~MeV (54.6 $\pm$ 2 $\%$). By considering all the four 
possibilities of transitions types of the 3$^{+}$ state (i.e. E$_1$ and M$_2$
for the 3$^{+}$ $\rightarrow$ 3$^{-}$ transition and, M$_1$ and E$_2$ for the 3$^{+}$ 
$\rightarrow$ 2$^{+}$ transition), our calculations yield to the conclusion that a value 
for the decay width $\Gamma_{\gamma}$ is fifty times lower than the one given in the 
literature \cite{Endt93,NNDC}, it means $\Gamma_{3^+}$ $<$ 0.23 eV. This 
result is important as it is the last known $\gamma$-decay level for the well 
studied $^{16}$O nucleus~\cite{Endt93,NNDC}. Much more details on the gamma-decay 
width estimates can be found elsewhere~\cite{Angel,Bormio}. 

There is a renew interest in the spectroscopy of the $^{16}$O nucleus at high 
excitation energy (i.e. larger than 12 MeV) as shown in Ref.~\cite{Ashwood09}, for 
instance. Charged particle data \cite{Freer95,Freer08} have been recently 
re-analysed to study higher excitation energy 4$\alpha$ states of the $^{16}$O 
nucleus near the $^{8}$Be+$^{8}$Be (14.619 MeV) and $^{12}$C+$\alpha$ (14.611 MeV) 
decay thresholds. In the framework of the study of Bose-Einstein Condensation (BEC) 
$\alpha$-particle state in light N=Z nuclei, the experimental signature of BEC in 
$^{16}$O is at present of the highest priority. An equivalent $\alpha$+''Hoyle" state 
in $^{12}$C is predicted to be the 0$^{+}_{6}$ state of  $^{16}$O at about 15.1 MeV, 
which energy is just lying (i.e. $\approx$ 700 keV) above the 4$\alpha$-particle 
breakup threshold \cite{Funaki08}. However, any state in $^{16}$O equivalent to
the so-called ''Hoyle" state in $^{12}$C is most certainly going to decay by
particle emission, with very small, probably un-measurable $\gamma$-decay
branches. Very efficient particle-detection techniques will have to be used in
the near future as such BEC states will be expected to decay by alpha emission
to the ''Hoyle" state, and could be associated with resonances in $\alpha$-particle
ineastic scattering on $^{12}$C leading to that state, or be observed in
$\alpha$-particle transfer to the $^{8}$Be-$^{8}$Be final state as observed
by P. Chevallier and coworkers back in the sixties \cite{Chevallier}. Another 
possibility might be to perform Coulomb excitation measurements with intense 
$^{16}$O beams at intermediate energies.

\newpage

\section{Conclusion}

The connection of $\alpha$-clustering, quasimolecular resonances, orbiting phenomena 
and extreme deformations (SD, HD, ...) has been discussed in this work by
using $\gamma$-ray spectroscopy of coincident binary fragments from either inelastic 
excitations and direct transfers (with small energy damping and spin transfer) or 
from orbiting (fully damped) processes \cite{Sanders99}. Exclusive data were 
collected with the Binary Reaction Spetrometer (BRS) in coincidence with 
{\sc EUROBALL~IV} installed formerly at the {\sc VIVITRON} Tandem facility of 
Strasbourg. The search for the $\gamma$-ray decay of a $^{12}$C+$^{12}$C molecule 
\cite{Greiner95,Erb85} as populated by projectile breakup has been undertaken. The 
existence of intense high-energy $\gamma$-rays in the E$_{x}$~=~10 MeV region in 
$^{24}$Mg$^{*}$ strongly populated in the $^{24}$Mg+$^{12}$C reaction is found to 
be of great interest. Most of these $\gamma$-rays may correspond to the doorway 
low-spin states recently observed in the radiative capture reaction 
$^{12}$C($^{12}$C,$\gamma$)$^{24}$Mg$^{*}$ \cite{Jenkins07,Jenkins05}.  

The striking selectivity of natural parity states in the $^{24}$Mg+$^{12}$C 
exit-channel indicates that the orbiting processes dominant in the vicinity
of the Coulomb barrier~\cite{Glaesner90} still survives at 
E$_{lab}$ = 130 MeV. From a careful analysis of the $^{16}$O+$^{20}$Ne 
$\alpha$-transfer exit-channel (strongly populated by orbiting) new information 
has been deduced on branching ratios of the decay of the 3$^{+}$ state of $^{16}$O 
at 11.089~MeV. This result is encouraging for a complete $\gamma$-ray spectroscopy 
of the $^{16}$O nucleus at high excitation energy. Of particular interest is the 
quest for the corresponding 4$\alpha$ states near the $^{8}$Be+$^{8}$Be and 
$^{12}$C+$\alpha$ decay thresholds.

The occurence of the collinear ternary decay from hyperdeformed shapes of 
$^{56}$Ni and $^{60}$Zn nuclei has been also investigated for $^{36}$Ar in the 
present work, but could not be firmly confirmed. However, this hypothesis in 
accordance with preliminary calculations performed in the framework of the 
generalized liquid-drop model \cite{Royer07} will need further 
experimental investigations with higher selectivity. The search for extremely 
elongated configurations (HD) in rapidly rotating medium-mass nuclei, which has 
been pursued exclusively using $\gamma$-ray spectroscopy, will have to be performed 
in conjunction with charged particle spectroscopy in the near future 
(see \cite{Oertzen08c,Wheldon08}).

\newpage

\noindent
{\small
{\bf Acknowledgments:} We thank both the staff of the {\sc VIVITRON} for providing 
us with good stable $^{24}$Mg beams and the EUROBALL group of Strasbourg for the 
excellent support in carrying out the experiment. Prof. Z. Dombradi and Prof. E. Uegaki
are acknowledged for their carefull reading of the mansucript. This work was supported 
by the french IN2P3/CNRS, the german ministry of research (BMBF grant under contract 
Nr.06-OB-900), and the EC Euroviv contract HPRI-CT-1999-00078. S.T. would like to 
express his gratitude for the warm hospitality during his three month stay in 
Strasbourg to the IReS and, he is grateful for the financial support obtained from 
the IN2P3, France.}

\end{document}